\documentclass{IEEEtaes}

\usepackage{array,amsthm}
\usepackage{amsmath,amssymb,amsfonts}
\usepackage{bm}
\usepackage{graphicx}
\usepackage{xcolor}

\usepackage[linesnumbered,ruled,vlined]{algorithm2e} 

\usepackage{tikz}
\usetikzlibrary{calc}
\usepackage{pgfplots}
\usepgfplotslibrary{statistics}
\pgfplotsset{compat=newest}
\pgfplotsset{plot coordinates/math parser=false} 

\jvol{XX}
\jnum{XX}
\jmonth{XXXXX}
\paper{1234567}
\pubyear{2022}
\doiinfo{TAES.2022.Doi Number}

\DeclareMathAlphabet{\mathpzc}{OT1}{pzc}{m}{n}

\newcommand{\sS}[0]{\mathcal{S}}
\newcommand{\sE}[0]{\mathcal{E}}
\newcommand{\sQ}[0]{\mathcal{Q}}
\newcommand{\sR}[0]{\mathcal{R}}
\newcommand{\sU}[0]{\mathcal{U}}

\newcommand{\SNn}[0]{\mathrm{SN}_0}

\newcommand{\cs}[0]{\mathpzc{s}}
\newcommand{\cS}[0]{\mathpzc{S}}
\newcommand{\cSstar}[0]{\mathpzc{S}^\star}
\newcommand{\csp}[0]{\mathpzc{s}^1}
\newcommand{\cSp}[0]{\mathpzc{S}^1}
\newcommand{\cSpsearch}[0]{\mathpzc{S}^{1,s}}
\newcommand{\cSpadd}[0]{\mathpzc{S}^{1,a}}
\newcommand{\csdp}[0]{\mathpzc{s}^2}
\newcommand{\cSdp}[0]{\mathpzc{S}^2}
\newcommand{\csinit}[0]{\mathpzc{s}^{\mathrm{int}}}
\newcommand{\cscur}[0]{\mathpzc{s}^{\mathrm{c}}}

\newcommand{\FOS}[0]{\mathrm{FOS}}
\newcommand{\MU}[0]{\mathrm{MU}}
\newcommand{\sortMU}[0]{\mathrm{sortMU}}

\DeclareMathOperator*{\argmax}{argmax}

\newcommand{\Break}{\textbf{break}}

\def\arXivPrint{1}  

\makeatletter
\newcommand\gobblestar
    {\futurelet\@let@token\@dogobblestar}

\def\@dogobblestar
    {\let\next=\relax
     \ifx\@let@token*%
        \def\next{\@gobble}%
     \else\ifx\@let@token[%
         \def\next{\@gobbleoptional}%
     \fi\fi
     \next}

\def\@gobbleoptional[#1]{}
\makeatother

\newcommand\doSingleLine[1]
    {\let\normalnewline=\\
     \let\\=\gobblestar
     #1%
     \let\\=\normalnewline}
\begin{document}

\title{Split-Aperture Phased Array Radar Resource Management for Tracking Tasks}

\author{Pepijn B. Cox}
\member{Member, IEEE}
\affil{Radar Technology Department, TNO, The Hague, The Netherlands} 

\author{Wim L. van Rossum}
\member{Member, IEEE}
\affil{Radar Technology Department, TNO, The Hague, The Netherlands} 



\receiveddate{
This project has received funding from the European Union's Preparatory Action on Defence Research under grant agreement No. 882407. }

\corresp{{\itshape (Corresponding author: P.B. Cox)}}

\authoraddress{Pepijn B. Cox and Wim L. van Rossum are with the Radar Technology Department of TNO, P.O. Box 96864, 2509 JG The Hague, The Netherlands (e-mail: \href{mailto:pepijn.cox@tno.nl}{pepijn.cox@tno.nl}, \href{mailto:wim.vanrossum@tno.nl}{wim.vanrossum@tno.nl}).}


%
%
\ifx\arXivPrint\undefined\else

\makeatletter
\twocolumn[{
\vspace{2cm}
This paper has been accepted for publication in the

\vspace{1cm}
\centerline{\textbf{\huge{IEEE Transactions on Aerospace and Electronic Systems}}}

\vspace{5cm}

\vspace{1cm}
\textbf{Citation}\\
P.B. Cox and W.L. van Rossum, ``\doSingleLine{\@title},'' in \textit{IEEE Transactions on Aerospace and Electronic Systems}, pp --, vol. --, no. --, 2025.

\vspace{1cm}

\definecolor{commentcolor}{gray}{0.9}
\newcommand{\commentbox}[1] {\colorbox{commentcolor}{\parbox{\linewidth}{#1}}}

\vspace{4cm}
\commentbox{
	\vspace*{0.2cm}
	\hspace*{0.2cm}More papers from P.B. Cox can be found at\\~\\
	\centerline{\large{\url{https://orcid.org/0000-0002-8220-7050}}}\\~\\
	\hspace*{0.2cm}and of W.L. van Rossum at\\
	\centerline{\large{\url{https://scholar.google.com/citations?user=Lh1u0qMAAAAJ}}}\\~\\
	\vspace*{0.2cm}
}

\vspace{3cm}
\textcopyright 2025 IEEE. Personal use of this material is permitted. Permission from IEEE must be obtained for all other uses, in any current or future media, including reprinting/republishing this material for advertising or promotional purposes, creating new collective works, for resale or redistribution to servers or lists, or reuse of any copyrighted component of this work in other works.
}]
\clearpage
\makeatother

\fi
%
%

\markboth{COX ET AL.}{Split-Aperture Phased Array Radar Resource Management}
\maketitle

\begin{abstract}
The next generation of radar systems will include advanced digital front-end technology in the apertures allowing for spatially subdividing radar tasks over the array, the so-called \emph{split-aperture phased array} (SAPA) concept. The goal of this paper is to introduce radar resource management for the SAPA concept and to demonstrate the added benefit of the SAPA concept for active tracking tasks. To do so, the radar resource management problem is formulated and solved by employing the \emph{quality of service based resource allocation model} (Q-RAM) framework. As active tracking tasks may be scheduled simultaneously, the resource allocation of tasks becomes dependent on the other tasks. The solution to the resource allocation problem is obtained by introducing the adaptive fast traversal algorithm combined with a three dimensional strip packing algorithm to handle task dependencies. It will be demonstrated by a simulation example that the SAPA concept can significantly increase the number of active tracks of a multifunction radar system compared to scheduling tasks sequentially.
\end{abstract}


\section{\uppercase{Introduction}}
F{\scshape uture} multifunction radar systems need to become more agile in contested and/or congested environments in complex operational theaters~\cite{Heras2022, Fernandez2024}. Agility can be obtained by integrating more advanced digital front-end technology into the phased array apertures. This front-end technology enables increased performance of radar tasks using multiple receive channels or using multiple transmit and multiple receive channels of the array, i.e., \emph{multiple-inputs multiple-outputs} (MIMO), concepts~\cite{Melvin2013,Bergin2018}.

Resource management for MIMO arrays using multi-beam concepts have been explored in, e.g.,~\cite{Yan2015a,Yan2015b}. For this multi-beam concept, the authors assume that it is possible to simultaneous transmit different waveform (tasks) in different angular directions with varying power between tasks while retaining sufficient suppression between the tasks and retaining constant modules of the signal to each transmit element in the array. However, solving this combined optimization problem remains an open challenge in the literature.

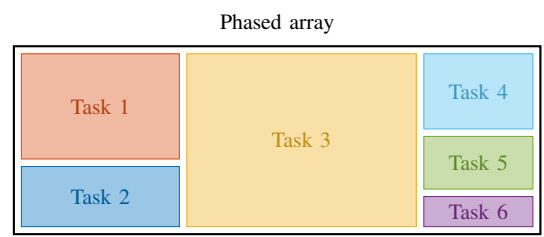
\begin{figure}[!b]%
\centering%
\definecolor{mycolor1}{rgb}{0.00000,0.44706,0.74118}%
\definecolor{mycolor2}{rgb}{0.85098,0.32549,0.09804}%
\definecolor{mycolor3}{rgb}{0.92900,0.69400,0.12500}%
\definecolor{mycolor4}{rgb}{0.49400,0.18400,0.55600}%
\definecolor{mycolor5}{rgb}{0.46600,0.67400,0.1880}%
\definecolor{mycolor6}{rgb}{0.30100,0.74500,0.9330}%
\definecolor{mycolor7}{rgb}{0.63500,0.07800,0.1840}%
\begin{tikzpicture}[font=\footnotesize]
	\draw [draw=black, thick] (0,0) rectangle ++(7,2.5);
	\node at (3.5,2.5) [above=1] {Phased array};
	\draw [draw=mycolor1, fill=mycolor1!40!white] (0.1,0.1) rectangle ++(2.10,0.8) node[pos=.5] {{\color{mycolor1!80!black}Task 2}};
	\draw [draw=mycolor2, fill=mycolor2!40!white] (0.1,1.0) rectangle ++(2.10,1.4) node[pos=.5] {{\color{mycolor2!80!black}Task 1}};
	\draw [draw=mycolor3, fill=mycolor3!40!white] (2.30,0.1) rectangle ++(3.05,2.3) node[pos=.5] {{\color{mycolor3!80!black}Task 3}};
	\draw [draw=mycolor4, fill=mycolor4!40!white] (5.45,0.1) rectangle ++(1.45,0.4) node[pos=.5] {{\color{mycolor4!80!black}Task 6}};
	\draw [draw=mycolor5, fill=mycolor5!40!white] (5.45,0.6) rectangle ++(1.45,0.7) node[pos=.5] {{\color{mycolor5!80!black}Task 5}};
	\draw [draw=mycolor6, fill=mycolor6!40!white] (5.45,1.4) rectangle ++(1.45,1.0) node[pos=.5] {{\color{mycolor6!80!black}Task 4}};
\end{tikzpicture}%
\caption{Illustration of the split-aperture phased array concept for simultaneously executing six tasks spatially divided over the array.}%
\label{fig:SAPA concept}%
\end{figure}

Alternatively, different tasks could be to simultaneously executed by spatially subdividing the array into subarrays, the so-called \emph{split-aperture phased array} (SAPA) concept. Figure~\ref{fig:SAPA concept} shows an example of dividing six tasks over the array. For each sub-array, modern, well-understood techniques for waveform optimization, beam forming, etc. can be used to execute the individual tasks. To demonstrate the effectiveness of the SAPA concept, in this paper, multiple active tracking tasks will be subdivided. The active tracking tasks can be subdivided by exploiting that the power-aperture budget of a radar system is usually designed for a scenario with a small, fast maneuvering target at a large distance. In cases with larger targets, slower maneuvering targets, and/or targets nearer to the radar, the transmit time and/or track update frequency is reduced to avoid overperformance on the tracking task. However, the transmit time and/or update frequency is lower bounded to maintain certain track requirements, e.g., Doppler resolution or robustness against unexpected target maneuvers. The remaining power-aperture budget in this minimal mode is still be excessive in certain cases.

Assigning radar resources to the active tracking tasks should be done adaptively to anticipate on the dynamic operational scene. Radar resource allocation has a vast literature, e.g., see~\cite{Miranda2006,Miranda2007,Moo2015,Greco2018,Charlish2017,Charlish2020,Yan2022} to name a few. The focus is on the optimization of active tracking tasks given the limited radar resources using the \emph{quality of service based resource allocation model} (Q-RAM) framework~\cite{Rajkumar1997}. The quality of the active tracking task is modeled by the \emph{Van Keuk and Blackman strategy} (KBS)~\cite{VanKeuk1993}. The combination of Q-RAM and KBS has a significant lower computational complexity to, e.g., neural-networks, entropy, nonlinear optimization, or dynamic programming based algorithms combined with other performance measures~\cite{Moo2015,Charlish2017,Yan2022}. The Q-RAM framework has been applied in different radar concepts including rotating radars~\cite{Yang2023}, multiple radar resources allocation problems~\cite{Irci2010}, and networks of radar systems~\cite{Nadjiasngar2015,Charlish2015}.

For the SAPA concept, the resource of a task, i.e., radar time budget, becomes coupled to other tasks. Scheduling concurrent tasks can be solved using a Monte Carlo tree search algorithm~\cite{Marquardt2024}. However, in this paper, also the loading of the tasks on the array is considered in the three dimensions: array in horizontal direction, array in vertical direction, and the combined radar time budget. Actually, optimally dividing the tasks over the array in time can be seen as an \emph{3 dimensional strip packing} (3SP) problem, i.e., determine an overlapping-free packing of the rectangles into a rectangle with fixed area and minimizing its height. 

Therefore, the SAPA resource allocation problem becomes a combination of maximizing the quality of the tasks given the radar resources that should be optimally scheduled via an 3SP like optimization problem. The 3SP is known to be NP-hard in the strong sense~\cite{Wascher2007} making it computationally challenging to find a solution. Efficiently solving the 3SP problem remains an active field of research. There exist many algorithms to solve the 3SP problem, e.g., via tree search~\cite{Fekete2007}, mixed-integer linear programming~\cite{Hifi2010}, or heuristic algorithms~\cite{Wauters2013,Christensen2017} to name a few. Generally, heuristic 3SP algorithms can significantly reduce the computational overhead but might lead to suboptimal solutions. As will be shown, the 3SP algorithm is executed many times in the resource allocation algorithm and therefore, the focus is on the heuristic 3SP algorithm~\cite{Wauters2013} with a relatively low computational overhead.

Obtaining a solution for the Q-RAM framework based resource allocation is known to have a large computational overhead. Especially, when the number of the control variables is large, as is the case for the SAPA resource allocation problem due to task coupling. To alleviate the computational burden, algorithms are introduced, such as the fast traversal algorithms~\cite{Hansen2004,Ghosh2006}, \emph{continuous double auction parameter selection} (CDAPS) algorithm~\cite{Charlish2015a}, or reinforcement learning~\cite{Durst2021} to approximate a solution. Approximation algorithms or the CDAPS algorithm can (significantly) reduce the explored input space compared to reinforcement learning by relying on the property that the marginal utility is monotonically increasing. For reinforcement learning, how to efficiently bound the explored input space by taking advantage of the monotonically increasing property for resource management is an open question. However, due to the task coupling in the SAPA concept, the combined resource function contains many discrete jumps within the discrete control space. The fast traversal algorithms and CDAPS algorithm cannot computationally efficient handle these cases. Therefore, in this paper, the adaptive fast traversal algorithm is introduced based on the fast traversal algorithm which can handle discrete jumps of the resource function in an adaptive manner.

The goal of this work is to define a radar resource management algorithm for the SAPA concept and demonstrate the added benefit for active tracking tasks. In particular, the Q-RAM framework is applied for resource management using the KBS model for active tracking combined with a 3SP algorithm to quantify the performance and resource measures. The resource allocation is solved by introducing the adaptive fast traversal algorithm. This paper extends our preliminary work~\cite{Cox2024} by
\begin{itemize}
	\item including the resource coupling between active tracking tasks, and
	\item introducing the adaptive fast traversal algorithm and combine it with a 3SP algorithm to solve the coupled SAPA resource allocation problem.
\end{itemize}

The paper is organized as follows. The resource management based on the Q-RAM framework is discussed in Section~\ref{sec:utility based resource management}. Section~\ref{sec:active tracking model} describes the active tracking performance and resource model. The coupled split-aperture resource model and 3SP algorithm is highlighted in Section~\ref{sec:split aperture resource model}. Solving the SAPA resource management is highlighted in Section~\ref{sec:solving RRM} where also the adaptive fast traversal algorithm is introduced. In Section~\ref{sec:example} the effectiveness of the SAPA concept is demonstrated by a simulation example followed by the conclusions in Section~\ref{sec:conclusions}.

\section{\uppercase{Utility Based Resource Management}} \label{sec:utility based resource management}

The utility based radar resource management problem is discussed in this section. 
The objective of the utility based radar resource management problem is to find the optimal control parameter selection $s=\{s_{t,1},…,s_{t,K} \}\in\sS$ given a set of tasks $\mathcal{K}$ at time $t$ to maximize the total utility~\cite{Charlish2017,Charlish2020,Yan2022}, as
\begin{subequations} \label{eq:constraint optimization RRM}
\begin{align}
\max_{s_t} u(s_t)&=\sum_k^K\omega_ku_k\left(q_k\left(s_{t,k},e_{t,k}\right)\right), \\
\mbox{s.t.}&\sum_k^Kg_k\left(s_{t,k},e_{t,k}\right)-r_{tot}\leq 0,
\end{align}
\end{subequations}
where $q_k:\sS_k\times\sE_k\rightarrow\sQ_k$ maps the control parameter space $s_{t,k}\in\sS_k$ and the environment space $e_{t,k}\in\sE_k$ at time $t$ into the quality space for a given task $k$. The resource function of the $k$-th task is $g_k:\sS_k\times\sE_k\rightarrow\sR_k$ and the total resource is bounded by $r_{tot}\in\mathbb{R}^+$. Then, $u_k:\sQ_k\rightarrow\sU=\left[0,~1\right]$ denotes the utility function of the $k$-th task with task weight $\omega_k\in\left[0,~1\right]$ normalized to $\sum \omega_k =1$.

Formulation~\eqref{eq:constraint optimization RRM} allows to allocate resources in an adaptive manner by obtaining a solution for a certain time window and solve it repetitively in time to adjust to the current environment and/or mission specific requirements. Mission specific requirements are included by, e.g., selection of weights $\omega_k$, type of utility functions $u_k$, etc. However, systematic translation of these requirements into~\eqref{eq:constraint optimization RRM} is still an open research topic and it requires deep expert knowledge.

For notational simplicity, the input arguments to the quality function $q_k$, the resource function $g_k$, and the utility function $u_k$ will be omitted in subsequent sections.

\section{\uppercase{Active Tracking Resource Allocation Model}} \label{sec:active tracking model}

The resource allocation model for the active tracking tasks are defined for the SAPA concept in this section. The model is based on the Van Keuk and Blackman strategy~\cite{VanKeuk1993} that is originally used to adaptively select the revisit interval for a tracking task. The KBS is an empirically determined function to maintain a certain maximum major axis of the posterior track covariance matrix, i.e., angular estimation error in u-v space. The strategy is based on assuming a Swerling I target amplitude fluctuation model with well-separated targets in space and it takes into account a complete tracking loop, including beam position over time, data association, and tracker dynamics based on a Singer model.


\subsection{Control Space}

The control parameters space $s_{t,k}=\{T_{d,k}, f_{t,k}, N_{h,k}, N_{v,k}\}$ consists of the coherent integration time $T_{d,k}\in\mathbb{R}^+$, the update frequency $f_{t,k}\in\mathbb{R}^+$, and the number of elements $N_{h,k},N_{v,k}\in\mathbb{N}$ used in the horizontal and vertical direction, respectively, at time $t$ for the $k$-th task. Note that, the parameters defining the control parameters space can be different per active tacking task.

\subsection{Environmental Space}

The environmental space $e_{t,k}=\{R_k,\theta_{h,k},\theta_{v,k},\sigma_k,\Theta_k,\Sigma_k\}$ consists of the target range $R_k\in\mathbb{R}^+$, the angle $\theta_{h,k},\theta_{v,k}\in\mathbb{R}$ in the horizontal and vertical direction, respectively, the radar cross section $\sigma_k\in\mathbb{R}^+$, and the standard deviation and time correlation $(\Theta_k, \Sigma_k)\in\mathbb{R}^+\times\mathbb{R}^+$ of Singer movement model at time $t$ for the $k$-th target. Standard tracking techniques can estimate the parameters in the environmental space $e_{t,k}$.

\subsection{Quality Function}

For the $k$-th active tracking task, the quality is defined by the angular estimation error~\cite{VanKeuk1993}
\begin{equation}
q_k = \theta_{bw}v_0,
\label{eq:quality active track}
\end{equation}
where $v_0\in\mathbb{R}^+$ denotes the track-sharpness, $\theta_{bw}=\max\left(\theta_{bw,h},\theta_{bw,v}\right)$ is the maximum of the half beamwidth in horizontal $\theta_{bw,h}$ or vertical $\theta_{bw,v}$ direction where $\theta_{bw,h}=\frac{\theta_{bw0,h}}{\theta_{h,k}}$ and $\theta_{bw,v}=\frac{\theta_{bw0,v}}{\theta_{v,k}}$ and the half beamwidth at boresight is taken as $\theta_{bw0,h}=\frac{\alpha_{bw}}{N_{h,k}}$ and $\theta_{bw0,v}=\frac{\alpha_{bw}}{N_{v,k}}$ assuming half wavelength element spacing with beamwidth factor $\alpha_{bw}\in\mathbb{R}^+$. The track-sharpness $v_0$ is obtained by finding the root of~\cite{Charlish2015}
\begin{equation}
	1+\left(\frac{\beta}{2}+2\right)v_0^2-\alpha\beta v_0^{2.4}=0,
	\label{eq:track sharpness}
\end{equation}
with
\begin{subequations}
\begin{align}
	\alpha &= 0.4f_{t,k} \left( \frac{R_k\theta_{bw}\sqrt{\Sigma_k}}{\Theta_k} \right)^{0.4}, \label{eq:alpha} \\
	\beta  &= \xi \SNn - \ln P_{fa} \label{eq:beta},
\end{align}
\end{subequations}
where $\xi\in(0,~1]$ denotes a cross-talk loss function, $\SNn\in\mathbb{R}^+$ is the \emph{signal-to-noise ratio} (SNR) without angular pointing error, and $P_{fa}\in\mathbb{R}^+$ defines the false alarm rate. Note that, in contrast to~\cite{Charlish2015,Charlish2017}, the angular estimation error and not the track-sharpness is used as a quality function in~\eqref{eq:quality active track}, because the beamwidth varies when changing the number of horizontal elements $N_{h,k}$ and/or vertical elements $N_{v,k}$ in the control space. 

Spatially dividing tasks will come at the cost of cross-talk between tasks. Therefore, the following loss function is applied\footnote{The cross-talk loss function is chosen arbitrarily as modeling of the front-end cross-talk in the SAPA concept is an open problem.}
\begin{equation}
	\xi = 0.8+0.2\frac{N_{h,k}}{N_{hT}}\frac{N_{v,k}}{N_{vT}},
	\label{eq:cross-talk loss}
\end{equation}
where $N_{hT},N_{vT}\in\mathbb{N}^+$ denote the total number of antenna elements in the horizontal and vertical dimension, respectively. The $\SNn$ in~\eqref{eq:beta} is given by~\cite{Richards2010}\footnote{The antenna gain is $G\approx\frac{4\pi\eta_\alpha A \cos\left(\theta_{h,k}\right)\cos\left(\theta_{v,k}\right)}{\lambda^2}$~\cite{Richards2010} and the array area is $A=\frac{\lambda}{2}N_{h,k}\frac{\lambda}{2}N_{v,k}$, which leads to $G\approx\pi\eta_\alpha N_{h,k}N_{v,k}\cos\left(\theta_{h,k}\right)\cos\left(\theta_{v,k}\right)$.}
\begin{equation}
	\SNn = k_{rad} \frac{N^3_{h,k}N^3_{v,k}T_d\cos^2\left(\theta_{h,k}\right)\cos^2\left(\theta_{v,k}\right)\sigma_t }{R_k^4},
	\label{eq:SN0}
\end{equation}
where the constant $k_{rad}\in\mathbb{R}^+$ is
\begin{equation}
	k_{rad}=\frac{P_{avg}\lambda^2\eta_\alpha^2}{64\pi k_bT_0FL_s},
	\label{eq:}
\end{equation}
with $P_{avg}\in\mathbb{R}^+$ denoting the average transmit power per element, $\lambda\in\mathbb{R}^+$ is the wavelength, $\eta_\alpha\in\mathbb{R}^+$ is the aperture efficiency, $k_b\in\mathbb{R}^+$ is the Boltzmann’s constant, $T_0\in\mathbb{R}^+$ is the receive system noise temperature, $F\in\mathbb{R}^+$ is the noise figure, and $L_s\in\mathbb{R}^+$ denoting the system losses.

In our paper, the maximum value of $\SNn$ is limited to 40\,dB and the minimum value to 10\,dB. We assume that the measurement accuracy is lower bounded due to practical limitations as calibration errors and the track sharpness $v_0$ in~\eqref{eq:quality active track} using $\SNn=40$\,dB. If $\SNn$ is below 10\,dB, it is assumed that the target will not be detected by the CFAR detector and, hence, the quality $q_k$ cannot be a number.

\subsection{Single Task Resource Function}

The expected steady-state resource is~\cite{Charlish2015}:
\begin{equation}
	g_k = n_l T_{d,k}f_{t,k},
	\label{eq:local resource function}
\end{equation}
where $n_l\in\mathbb{R}^+$ defines the expected number of looks given by~\cite{VanKeuk1993}
\begin{subequations}
\begin{equation}
	n_l=\frac{1}{P_D} \left(1+\left(\gamma v_0^2\right)^2 \right)^{1/2},
	\label{eq:nr looks}
\end{equation}
with $\gamma\in\mathbb{R}^+$ and the probability of detection $P_D\in\mathbb{R}^+$, assuming a Swerling I target fluctuation model, are
\begin{align}
\gamma &\cong 1+14\left(\frac{\vert \ln P_{fa}\vert}{\xi\SNn}\right)^{1/2}, \label{eq:nr looks dist} \\
P_D 	 &= P_{fa}^{\frac{1}{1+\xi\SNn}}. \label{eq:nr looks Pd} 
\end{align}
\end{subequations}
Equivalent to the quality function $q_k$ in \eqref{eq:quality active track}, if the $\SNn$ is below 10\,dB, then the resource $g_k$ will not be a number. In cases with low $\SNn$, the resource function $g_k$ in~\eqref{eq:local resource function} accounts for the potential need of multiple observations of the target.

Note that the number of elements $N_{h,k}$ and $N_{v,k}$ have an impact on both the quality $q_k$ in \eqref{eq:quality active track} and on the resource function $g_k$ in \eqref{eq:local resource function} via, amongst others, the beamwidth $\theta_{bw}$, the $\SNn$, and the cross-talk loss function.

\subsection{Single Task Utility Function}

The utility function $u_k$ characterizes the level of satisfaction of each task. In this paper, a linear utility function is applied given by
\begin{equation}
	u_k = \max\left(\min\left(\frac{q_k-q_{k,min}}{q_{k,max}-q_{k,min}},1\right), 0\right),
	\label{eq:linear utility}
\end{equation}
where $q_{k,min},q_{k,max}\in\mathbb{R}$ are the minimum and maximum values for the angular estimation error for the $k$-th task. The $\min$ and $\max$ functions in the utility function~\eqref{eq:linear utility} are used to bound on the interval $[0,~1]$.

\section{\uppercase{Split-Aperture Resource Function}} \label{sec:split aperture resource model}

\begin{figure}%
\centering
	\includegraphics[width=0.7\columnwidth, angle=0, trim={1.2cm 0.3cm 2cm 1.6cm}, clip]{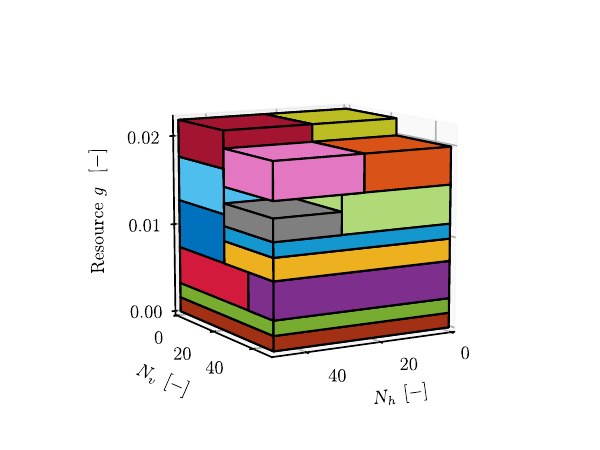}%
\caption{An illustrative example of distributing 15 tracking tasks over the array in the horizontal dimension $N_h$ and vertical dimension $N_v$ with total resource $g_k=0.0218$.}%
\label{fig:stacking}%
\end{figure}

The tracking tasks can be considered blocks in the number of horizontal elements $N_{h,k}$, the number of vertical elements $N_{v,k}$, and resource $g_k$. These blocks should be scheduled on the array along the resource dimension where all tasks cannot intersect. An illustrative example of distributing 15 tracking tasks is given in Figure~\ref{fig:stacking}. Due to these constrains, the tasks that are divided over the array become coupled. Actually, dividing the tasks over the array in time can be seen as an \emph{3 dimensional strip packing} (3SP) problem, i.e., determine an overlapping-free packing of the rectangles into a rectangle with fixed area and minimizing its height. The 3SP is known to be NP-hard in the strong sense~\cite{Wascher2007}.

For the radar resource allocation problem, the heuristic 3SP algorithm of~\cite{Wauters2013} is employed which has a relatively low computational overhead. The algorithm is highlighted next.

\subsection{3D Strip Packaging}

In this section, the heuristic strip packing algorithm of~\cite{Wauters2013} is highlighted. The technique is based on the \emph{deepest-bottom-left-fill} (DBLF) algorithm combined with a so-called shaking procedure. The DBLF algorithm highlighted in Algorithm~\ref{alg:DBLF} is known to lead to suboptimal solutions in some situations. The additional shaking technique in Algorithm~\ref{alg:shaking} is used improve the packing solution with low computational overhead. In this paper, the boxes that are placed cannot be rotated in contrast to the algorithm in~\cite{Wauters2013}, as this may change the quality of the task.

The DBLF method in Algorithm~\ref{alg:DBLF} is initialized with an ordered list of tasks $I$ and it constructs a packaging solution $S$. The DBLF method places the tasks in turn to the deepest available position, i.e., smallest $z$ coordinate, then as far to the bottom as possible, i.e., smallest $y$ coordinate and, finally, as far to the left as possible, i.e., smallest $x$ coordinate. The $\mathrm{getAllGaps}$ method provides an ordered list with the potential coordinates to place the boxes. The gaps are ordered according to deepest-bottom-left. In this paper, the \emph{extreme point} (EP) based heuristics of~\cite{Crainic2008} have been used which scales on $\mathcal{O}(n)$ where $n$ represents the number of boxes to stack. The $\mathrm{fits}$ methods verifies if the $i$-th item does not overlap with other boxes in the current solution $S$.


\begin{algorithm}[!t]
\DontPrintSemicolon
\caption{DBLF procedure}\label{alg:DBLF}
\KwIn{I}
$S \gets \emptyset$ \\
\For {$i \in I $}{
	$\mathrm{placed}\gets \mathrm{false}$ \\
	$\mathrm{gaps}\gets \mathrm{getAllGaps}(S)$ \\
	\For {$g \in \mathrm{gaps}$}{
		\uIf{$\mathrm{placed}=\mathrm{false}$ and $\mathrm{fits}(i,g,S)$}{
			remove $g$ from $\mathrm{gaps}$ \\
			add $i$ to $S$ in position $g$ \\
			$\mathrm{placed}\gets \mathrm{true}$ \\
		}
}}
\KwOut{S}
\end{algorithm}

The shaker technique in Algorithm~\ref{alg:shaking} iterates between a forward shake and a backward shake where it re-sorts the order of placing the object based on sorting criterion $c\in\mathcal{C}$. In addition, the shaking can be repeated multiple times by providing $k>1$.

The sorting criterion $c$ can be based on the $(x_k,y_k,z_k)$-coordinate of the $k$-th task, box size $N_{h,k}$ $N_{v,k}$ and $g_k$, volume, area, etc. or combinations thereof~\cite{Crainic2008,Wauters2013}. As suggested in~\cite{Wauters2013} in case of limited computational time, the following two sorting options can be used
\begin{align*}
c_1 & &z_k+g_k~;~y_k+N_{v,k}~;~x_k+N_{h,k}, \\
c_2 & &z_k+g_k~;~x_k+N_{h,k}~;~y_k+N_{v,k},
\end{align*}
where the ordering $z_k+g_k;y_k+N_{v,k};x_k+N_{h,k}$ implies that it is sorted by decreasing $z_k$-coordinate plus the single task resource $g_k$, ties are broken using decreasing $y_k$-coordinate plus the number of vertical elements $N_{v,k}$ and then $x_k$-coordinate plus the number of horizontal elements $N_{h,k}$.


\begin{algorithm}[!t]
\DontPrintSemicolon
\caption{Shaking procedure~\cite{Wauters2013}}\label{alg:shaking}
\KwIn{I, C, k}
$S \gets \mathrm{DBLF}(I)$\\
$S^\star \gets S$\\
$i \gets 0$\\
\While{$i < k $}{
	\For{$c \in C$}{
		$I' \gets \mathrm{sort}(S^\star,c)$ \\
		$S' \gets \mathrm{DBLF}(I')$ \\
		\uIf {$g(S')<g(S^\star)$}{
			$S^\star\gets S'$
		}
		$I'' \gets \mathrm{sort}(S',c)$ \\
		$S'' \gets \mathrm{DBLF}(I'')$ \\
		\uIf {$g(S'')<g(S^*)$}{
			$S^\star\gets S''$
		}
	}
	$i\gets i+1$
}
\KwOut{$S^\star$}
\end{algorithm}

\subsection{Split-Aperture Resource Function}

The split-aperture resource function for a set of tasks $\mathcal{K}$ is given by
\begin{equation}
g(S^\star) = \max_{k\in \mathcal{K}} z_k+g_k,
\label{eq:height functions}
\end{equation}
where the solution of the bin stacking problem $S^\star$ can be obtained by the shaking procedure in Algorithm~\ref{alg:shaking}.

\subsection{Split-Aperture Utility Function}

The split-aperture utility function $u$ for a set of tasks $\mathcal{K}$ is given by
\begin{equation}
u = \sum_{k\in \mathcal{K}} \omega_k u_k,
\label{eq:split-apreture utility function}
\end{equation}
to combine individual tasks.

\section{\uppercase{Solving SAPA Radar Resource Management}} \label{sec:solving RRM}

The Q-RAM formulation is well-known to solve the Karush-Kuhn-Tucker optimality conditions for a discrete control space $s_{t,k}$ under the assumptions that the utility $u_k$ and resource $g_k$ functions are concave functions in $s_{t,k}$ given $e_{t,k}$~\cite{Irci2010,Charlish2017}. Solving the optimization~\eqref{eq:constraint optimization RRM} can be performed by using different Q-RAM solvers~\cite{Rajkumar1997}, including convex hull approximation algorithm~\cite{Hansen2004,Ghosh2006}, CDAPS~\cite{Charlish2015a}, and reinforcement learning~\cite{Durst2021}. The basic algorithm~\cite{Rajkumar1997} to solve~\eqref{eq:constraint optimization RRM} has certain drawbacks. It achieves near-optimal solutions due to sub-optimal stopping conditions, as the optimal solution might not lie on the concave-majorant~\cite{Irci2010,Charlish2017}. To minimize sub-optimality, the discrete control space should contain sufficient granularity. On the other hand, the computational complexity can be high when the dimension of the discrete control parameter space is large. Especially, in the SAPA case with simultaneous task scheduling, the tasks become coupled via the resource function. In Section~\ref{sec:example}, the example defined leads to roughly $\sim3\cdot10^{298}$ set-points in the control parameter set for only 60 active tracking tasks. Hence, next, a traversal technique is derived to approximate the concave-majorant by a tree like search. Then, the resource allocation problem in~\eqref{eq:constraint optimization RRM} is solved using the following steps~\cite{Rajkumar1997}:
\begin{enumerate}
	\item Construct an approximation of the concave-majorant for all tasks using Algorithm~\ref{alg:AFT} and construct a list of all utility $u$ and resource $g$ set-points
	\item Order the set-points of the concave-majorant for all tasks in descending order based on the marginal utility. The marginal utility is the difference in utility divided by the difference in resource between set-points.
	\item Traverse over the sorted list with highest marginal utility to allocate resource until no resource remains.
\end{enumerate}

\subsection{Approximating the Concave-Majorant} \label{subsec:ApproxConcaveMajorant}

Constructing the concave-majorant by evaluating all set-points for all discrete control parameter in $\sS$ is not efficient or, in the SAPA case, impossible. In this section, the adaptive traversal algorithm is introduced to handle the discrete jumps in the concave-majorant approximation based on the fast traversal algorithm~\cite{Ghosh2006} with low computational complexity.

\begin{figure}[!t]%
\centering
\pgfdeclarelayer{back}%
\pgfsetlayers{back,main}%
\makeatletter%
\pgfkeys{%
  /tikz/on layer/.code={
    \def\tikz@path@do@at@end{\endpgfonlayer\endgroup\tikz@path@do@at@end}%
    \pgfonlayer{#1}\begingroup%
  }%
}%
\makeatother%
\definecolor{mycolor1}{rgb}{0.00000,0.44706,0.74118}%
\definecolor{mycolor2}{rgb}{0.85098,0.32549,0.09804}%
\definecolor{mycolor3}{rgb}{0.92900,0.69400,0.12500}%
\definecolor{mycolor4}{rgb}{0.49400,0.18400,0.55600}%
\definecolor{mycolor5}{rgb}{0.46600,0.67400,0.1880}%
\definecolor{mycolor6}{rgb}{0.30100,0.74500,0.9330}%
\definecolor{mycolor7}{rgb}{0.63500,0.07800,0.1840}%
\begin{tikzpicture}[font=\footnotesize]

\draw[thick,->] (0,0) -- (4.5,0) node[pos=0.5, anchor=north] {resource $g_k$};
\draw[thick,->] (0,0) -- (0,3) node[left, pos=0.5, rotate=90, anchor=south] {utility $u_k$};

\node (start) at (0.5,0.2)[circle,fill,inner sep=1.5pt]{};

\node (c1) at ($(start) + (0.8,0.25)$)[circle,fill, mycolor1,inner sep=1.5pt]{};
\node (c2) at ($(start) + (2.4,0.2)$)[circle,fill, mycolor1,inner sep=1.5pt]{};

\node (c11) at ($(c1) + (0.4,0.5)$)[circle,fill, mycolor2,inner sep=1.5pt]{};
\node (c12) at ($(c1) + (0.5,0.4)$)[circle,fill, mycolor2,inner sep=1.5pt]{};
\node (c13) at ($(c1) + (0.8,0.2)$)[circle,fill, mycolor2,inner sep=1.5pt]{};

\node (c21) at ($(c2) + (0,0.7)$)[circle,fill, mycolor2,inner sep=1.5pt]{};
\node (c22) at ($(c2) + (0,1.2)$)[circle,fill, mycolor2,inner sep=1.5pt]{};
\node (c23) at ($(c2) + (0.6,0.2)$)[circle,fill, mycolor2,inner sep=1.5pt]{};

\node (c211) at ($(c21) + (0,1.5)$)[circle,fill, mycolor3,inner sep=1.5pt]{};
\node (c211c) at (c211) [thick,circle,fill=none,draw=mycolor1, densely dotted,inner sep=2.5pt]{};
\node (c221) at ($(c22) + (0,0.3)$)[circle,fill, mycolor3,inner sep=1.5pt]{};

\node (c2111) at ($(c211) + (0.5,0.4)$)[circle,fill, mycolor4,inner sep=1.5pt]{};
\node (c2112) at ($(c211) + (0.6,0.3)$)[circle,fill, mycolor4,inner sep=1.5pt]{};
\node (c2113) at ($(c211) + (0.7,0.2)$)[circle,fill, mycolor4,inner sep=1.5pt]{};
\node (c2211) at ($(c221) + (0.6,0.3)$)[circle,fill, mycolor4,inner sep=1.5pt]{};
\node (c2212) at ($(c221) + (0.45,0.2)$)[circle,fill, mycolor4,inner sep=1.5pt]{};
\node (c2213) at ($(c221) + (0.3,0.1)$)[circle,fill, mycolor4,inner sep=1.5pt]{};

\draw[mycolor1,on layer=back] (start) -- (c1){};
\draw[mycolor1,on layer=back] (start) -- (c2){};

\draw[mycolor2,on layer=back] (c1) -- (c11){};
\draw[mycolor2,on layer=back] (c1) -- (c12){};
\draw[mycolor2,on layer=back] (c1) -- (c13){};

\draw[mycolor2] (c2) -- (c21) node[right, pos=1, anchor=west, black] {1};
\draw[mycolor2] (c2) -- (c22) node[right, pos=1, anchor=west, black] {2};
\draw[mycolor2] (c2) -- (c23) node[right, pos=1, anchor=north west, black] {3};

\draw[mycolor3,on layer=back] (c21) -- (c211) {} ;
\draw[mycolor3,on layer=back] (c22) -- (c221) {} ;

\node[anchor=south west, yshift=1mm] at (c211) {1};
\node[anchor=south west, yshift=1mm] at (c221) {2};

\draw[mycolor4,on layer=back] (c211) -- (c2111){};
\draw[mycolor4,on layer=back] (c211) -- (c2112){};
\draw[mycolor4,on layer=back] (c211) -- (c2113){};
\draw[mycolor4,on layer=back] (c221) -- (c2211){};
\draw[mycolor4,on layer=back] (c221) -- (c2212){};
\draw[mycolor4,on layer=back] (c221) -- (c2213){};

\path[thick,->,mycolor1] (c2) edge [bend left=20] node[left,xshift=1mm] {update} (c211c) ;

\draw[dashed,black,on layer=back] (start) -- (c211c){};

\end{tikzpicture}%
%
\caption{Adaptive fast traversal concept of updating the approximate concave-majorant including jumps. The black color represents the current point on the concave-majorant. Assuming control space with three variables. The blue color represent the first order set-point, the red indicates the second order set-point, yellow the third order, and magenta the fourth order.}%
\label{fig:resource jumps}%
\end{figure}

The fast traversal algorithms are based on the concept that the marginal utility, i.e., $u_k/g_k$, is monotonically increasing. The utility and resource functions are chosen to be monotonically increasing, which comes natural for resource management as an increasing value in the discrete control spaces $s_{t,k}$ should lead to more resources and an equal or increased utility.

Given a set of discrete control spaces $s_{t,k}$, the fast traversal algorithm obtains the next step on concave-majorant by keeping all index values of the current control space, denoted by $\cs\in\mathbb{N}^N$, equal but except for one. By varying the index value among all control dimensions will lead to all set-points for this step from which point is selected to maximize the marginal utility. Note that the marginal utility is monotonically increasing, hence, only positive index values need to be considered. The first-order next index value of the discrete control set $\sS$ is obtained by
\begin{equation}
 \FOS(\cs) =  \{ \cs+ \mathbf{1}_1, \ldots, \cs+ \mathbf{1}_N\},
\label{eq:FOS}
\end{equation}
where $\mathbf{1}_i$ denotes a vector with all zeros except for the $i$-th element that is one and $N=\sum_k \dim(\sS_k)$. With a slight abuse of notation, the first-order approximation of the marginal utility around the current point $\cscur$ with $\cS=\FOS(\cscur)$ is given by
\begin{equation}
 \MU(\cS,\cscur) =  \frac{u_k(\cS) - u_k(\cscur)}{g_k(\cS) - g_k(\cscur)}
\label{eq:marginal utility set}
\end{equation}
where $\MU(\cS,\cscur)\in\mathbb{R}^N$ is a vector for which the $i$-th element is $\left(u_k([\cS]_i) - u_k(\cscur)\right)/\left(g_k([\cS]_i) - g_k(\cscur)\right)$ with $[\cS]_i$ denoting the $i$-th index value out of the set $\cS$. Initializing with the minimal starting point $s^{\mathrm{init}}$ that takes boundary conditions into account\footnote{In our case, $\csinit$ should be selected such that the input space represents the point where the $\SNn$ exceeds $10$\,dB.}, the first-order traversal of~\cite{Hansen2004,Ghosh2006} iteratively selects the next point by selecting the maximum element out of $\MU(\cS,\cscur)$. 

Based on this concept, in this paper, the \emph{adaptive fast traversal} (AFT) in Algorithm~\ref{alg:AFT} is introduced to approximate concave-majorant that can handle discrete jumps in the marginal-utility due to the split-aperture resource allocation. The algorithm starts by updating the current index values point $\cscur$ to the initial set $\csinit$ and by adding the initial set $\csinit$ to the set of index value points that make up the approximate concave-majorant $\cSstar$. Then, the set of first-order index points are obtained by $\cSp=\FOS(\cs)$.


\begin{algorithm}[!t]
\DontPrintSemicolon
\caption{Adaptive fast traversal}\label{alg:AFT}
\KwIn{$\csinit$, $\alpha_1$, $n_1$, $n_2$, $n_3$}
$\cscur,\cSstar\gets\csinit$ \\
$\cSp,\cSpsearch\gets \FOS(\csinit)$ \\
\While {$u(\cscur) < 1 $}{
	$n\gets0$ \\
	\While {$\mathrm{true}$}{
		$i\gets \argmax \MU(\cSp,\cscur)$ \label{alg:max marginal utility} \\
		$k\gets0$ \\
		$\cSpadd\gets\emptyset$ \\
		\For {$\csp \in \sortMU(\cSpsearch,\cscur)$} {\label{alg:sort decending MU}
			$k\gets k+1$ \\
			\If {$\MU(\csp,\cscur)$ \newline 
					 \hspace*{1cm}$< \phi(n;\alpha_1,n_1) \MU([\cSp]_i,\cscur)$ \newline
					 \hspace*{2cm} or $k>n_2$} { \label{alg:if First Brach Stop} 
				 \Break
			}
			$\cSdp\gets \FOS(\csp)$ \\
			$m\gets0$ \\
			\For {$\csdp \in \sortMU(\cSdp,\cscur)$ \newline
					   \hspace*{1cm}  with $g(\csdp) == g(\csp)$} { \label{alg:second-order search}
				$m\gets m+1$ \\
				\If{$m>n_3$} {
					\Break
				}
				Add $\csdp$ to $\cSpadd$ \\
			}
		}
		\uIf {$\cSpadd$ is not empty} {\label{alg:update set}
			Add $\cSpadd$ to $\cSp$ \\
			$\cSpsearch\gets\cSpadd$ \\
			\If {all $\MU(\cSpadd,\cscur) < \MU([\cSp]_i,\cscur)$}{
				$n\gets n+1$ \label{alg:increment adaptivity} \\
			}
		}\Else{
			\Break
		}
	}
	$\cscur\gets [\cSp]_i$ \label{alg:update current point} \\
	Add $\cscur$ to $\cSstar$ \label{alg:update concave-majorant points} \\
	$\cSp\gets\FOS(\cscur)$ \\
}
\KwOut{$\mathrm{concaveMajorant}(\cSstar)$} \label{alg:return concave majorant}
\end{algorithm}

Algorithm~\ref{alg:AFT} will recursively evaluate if the second-order indexes $\cSdp$ potentially include an increased utility with equivalent resource, i.e., discrete step in the marginal utility. The adaptive fast traversal will lead to a tree like search of branches to search for potential discontinuities. To do so, the method $\sortMU$ sorts the first-order set $\cSpsearch$ based on the marginal utility in~\eqref{eq:marginal utility set} with respect to the current point $\cscur$ in descending order (line~\ref{alg:sort decending MU}~of Algorithm~\ref{alg:AFT}). 

In order to keep the number of searches in the second-order set to a minimum, the search is stopped if the marginal utility of the point $\csp$ is a fraction $\phi$ below the largest marginal utility in all points in the first-order set $\cSp$ (line~\ref{alg:if First Brach Stop}) with
\begin{equation}
\phi(n;\alpha_1,n_1) = \left(1-\alpha_1\right)\frac{n}{n_1}+\alpha_1,
\label{eq:function MU increments}
\end{equation}
or when the number of iterations $k$ is larger than $n_2$. In case that the second-order search does not lead to improved marginal utility with points that have equivalent resource, then the fraction $\phi$ in~\eqref{eq:function MU increments} will stepwise go to one and the method adaptively reduces the number of tree branches searched (by incrementing $n$ in line~\ref{alg:increment adaptivity}).

Then, the second-order set $\cSdp=\FOS(\csp)$ is recursively evaluated in descending order based on the marginal utility for points where the resource function between the first-order point $\csp$ and second-order point $\csdp$ is equal (line~\ref{alg:second-order search}). In such case, the second-order point $\csdp$ is added to the first-order set $\cSpadd$ and only a maximum of $n_3$ index value points will be considered. The auxiliary variables $\cSpsearch$ and $\cSpadd$ are necessary to track the current index value points to be searched $\cSpsearch$ for potential discrete steps and the set of points $\cSpadd$ to be added to $\cSp$.

After iterating on the set $\cSpsearch$, potentially new points could have been found which are stored in $\cSpadd$ (lines~\ref{alg:update set}). The set of index value points $\cSp$ and $\cSpsearch$ are updated and a new iteration of the adaptive search is started until $\cSpadd$ remains empty. In this case, all index value points in $\cSp$ with large marginal utility have a guaranteed next point without discontinuous marginal utility. Then, the index value point with maximum marginal utility becomes the next index value point $\cscur$ (line~\ref{alg:update current point}) and this index value point is added to set of index values to create the approximate concave-majorant $\cSstar$ (line~\ref{alg:update current point}). The algorithm will execute until the total utility of the current point $\cscur$ reaches one, i.e., $u(\cscur)=1$, as then all tasks have reached maximum utility. Finally, some points in $\cSstar$ might not be on the concave-majorant and, hence, the convex hull of $\cSstar$ should be constructed (line~\ref{alg:return concave majorant})~\cite{Graham1972}.

The tree like search in Algorithm~\ref{alg:AFT} is adaptive. By increasing $n_1$, $n_2$, $n_3$ or decreasing $\alpha_1$, the AFT method explores more potential candidates that could have an increased marginal utility. On the other hand, considering more index value points implies that the quality function, utility function, and, most prominent, the SAPA resource function need to be evaluated. This may significantly increase the computational overhead.

Algorithm~\ref{alg:AFT} can be extended with efficient booking to reduce the computational load. For example, the evaluation of the quality function, utility function, and the resource function for the explored index value points should be stored to avoid re-computation. In step $l$, these functions are evaluated on line~\ref{alg:second-order search} and these could be re-used in step $l+1$ for lines~\ref{alg:max marginal utility}, \ref{alg:sort decending MU}, and \ref{alg:if First Brach Stop}. Also, when a certain task $k$ reaches local utility $u_k$ equal to one then the index values corresponding to this task should not be increased anymore in the $\FOS$ function~\eqref{eq:FOS}.

\section{\uppercase{Split-Aperture Resource Allocation Example}} \label{sec:example}

In this section, the split-aperture resource allocation problem of the SAPA concept is compared to the resource allocation where the radar cannot split the aperture to demonstrate the benefits of the SAPA concept. In Section~\ref{subsec:simulation setting}, the simulation setting is provided and the results are discussed in Section~\ref{subsec:Simulation results}.

\subsection{Simulation Setting} \label{subsec:simulation setting}

\begin{table}[!t]
\caption{The parameter ranges of the Singer movement model~\cite{Charlish2015a}.} 
\label{tbl:Singer Model}
\centering
\begin{tabular}{|c|c|c|} 
 \hline
  & $\Theta_{t,k}~[\text{\,m/s}^{\text{2}}]$ & $\Sigma_{t,k}~[\text{s}]$  \\
 \hline\hline
 Type I & 20-35 & 10-20 \\ \hline
 Type II & 0-5 & 1-4 \\ \hline
 Type III & 5-20 & 30-50 \\ \hline
\end{tabular}
\end{table}

The simulation setting for the split-aperture resource allocation example is discussed in this section, which includes the scene, the target parameters, the radar parameters, and the resource allocation parameters. The scene will contain 60 targets where the environmental space $e_{t,k}$ of the $k$-th target is defined as follows. The horizontal and vertical angles are a realization from a uniform distribution on $\theta_{h,k}\sim \mathpzc{U}(-60,60)^{\circ}$, $\theta_{v,k}\sim\mathpzc{U}(0,70)^{\circ}$, and the RCS on $\sigma_k\sim\mathpzc{U}(-10,10)\text{\,dBm}^{\text{2}}$. Two scenes will be simulated. In the first scene, the targets are placed randomly in range according to $R_k\sim\mathpzc{U}(10,70)$\,km and, in the second scene, according to $R_k\sim\mathpzc{U}(10,250)$\,km. If the target is above an altitude of 20\,km then the vertical angle $\theta_{v,k}$ is changed based on the drawn target range $R_k$ with an altitude that is resampled from a uniform distribution $\mathpzc{U}(0,20)$\,km. For the singer model parameters, the targets have an equal chance in being type I, type II or type III and, after the type is randomly selected, the standard deviation $\Theta_k$ and time correlation $\Sigma_k$ are randomly drawn from a uniform distribution with the domain as indicated in Table~\ref{tbl:Singer Model}. In radar operations, some targets have higher priority than others. To demonstrate this, twelve targets will have a high priority with a task weight uniformly drawn from $\omega_k\sim\mathpzc{U}(0.7,0.9)$ and the remaining targets have a task weight uniformly drawn from $\omega_k\sim\mathpzc{U}(0.2,0.5)$.

For the radar, the following constant is used $k_{rad}=2.4\cdot 10^{16}\text{\,m}^\text{2}/\text{s}$, the false alarm rate is set at $P_{fa}=10^{-4}$, the radar tilt angle is $5^{\circ}$, and the total amount of elements in the horizontal and vertical dimensions are $N_{hT}=N_{vT}=48$.

For the control space of each tracking task, the coherent integration time $T_{d,k}$ can be chosen from the discrete set $[4,~5.2,\ldots,64]$\,ms, the update rate $f_{t,k}$ from $[0.2,~0.4,\ldots,6]$\,Hz, and the number of horizontal elements $N_{h,k}$ and vertical elements $N_{v,k}$ from $[6, 12,\ldots,48]$. Note that the split-aperture allocation problem also includes $N_{h,k}=N_{v,k}=48$ elements in the control space, i.e., it can assign the full array for a certain task if necessary. The utility function has as its minimal value $q_{k,min}=3$\,mrad and its maximum value $q_{k,max}=1$\,mrad, which implies this $u_k=1$ if $q_k\leq 1$\,mrad and $u_k=0$ if $q_k\geq3$\,mrad. For the simulation example, $N_{MC}=100$ Monte Carlo runs are conducted where each run has a new realization of the environmental space $e_{t,k}$ and weights $\omega_k$.

The shaking procedure is initialized with an ordering of tasks descending on area on the array, i.e., $N_{h,k}N_{v,k}$ and with ties are broken in descending on the local resource $g_k$. The shaking procedure will take $k=1$ steps. For the AFT algorithm, the parameters $\alpha_1=0.7$, $n_1=2$, and $n_2=n_3=3$ are selected.

The introduced resource SAPA allocation problem includes constrains on scheduling the tasks on the array via the 3SP problem. To asses the performance, it is compared to the case that all tasks can only be scheduled on the full-aperture and the case without constrains on scheduling the tasks on the array. The resource allocation without contains can be seen as the lower bound on the resource and the full-aperture case can be considered as the upper bound on the resource. Note that, in an actual system, the constrains need to be considered. The full-aperture case is considered by using only $N_{h,k}=N_{v,k}=48$ elements in the horizontal and vertical dimension. For the SAPA resource allocation without constrains, the individual resource functions are taken as~\cite{Cox2024}
\begin{equation*}
	g_{u,k} = n_l T_{d,k}f_{t,k}\frac{N_{h,k}N_{v,k}}{N_{hT}N_{vT}},
\end{equation*}
where the task resource function~\eqref{eq:local resource function} has been changed to include the ratio of the loading on the horizontal and vertical dimension, i.e., by adding $\frac{N_{h,k}N_{v,k}}{N_{hT}N_{vT}}$. Note that for the unconstrained SAPA case and the full-aperture case, the individual tasks become decoupled and first-order fast traversal technique can be used to obtain the concave-majorant for the individual tasks.

\begin{figure}%
\input{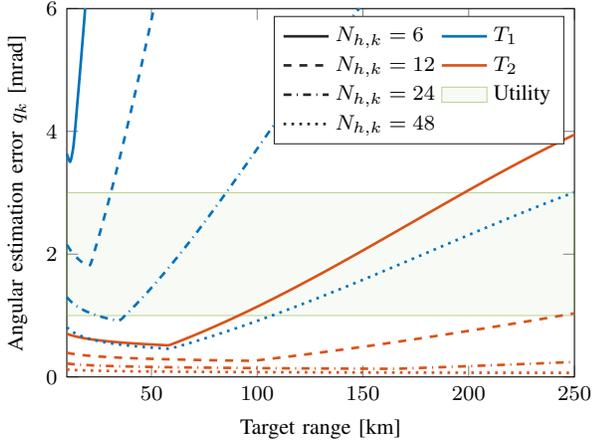}%
\caption{The quality function of the active track task $q_k$. The angular estimation error is displayed for different target ranges, number of horizontal elements $N_{h,k}$, and target parameters $T_1$,$T_2$ for an update rate of $f_{t,k}  = 2$\,Hz.}%
\label{fig:quality function}%
\end{figure}

Figure~\ref{fig:quality function} shows the quality function $q_k$ of the $k$-th active tracking task, i.e., the angular estimation error. The figure shows target $T_1$ with parameters $\sigma_k=0.1\text{\,m}^{\text{2}}$, $T_{d,k}=64$\,ms, $\Sigma_k=35\text{\,m/s}^{\text{2}}$, $\Theta_k=10$\,s, $\theta_{h,k}=\theta_{v,k}=60^\circ$, and target $T_2$ with parameters $\sigma_k=10\text{\,m}^{\text{2}}$, $T_{d,k}=20$\,ms, $\Sigma_k=0.1\text{\,m/s}^{\text{2}}$, $\Theta_k=4$\,s, $\theta_{h,k}=0^{\circ}$, $\theta_{v,k}=0^{\circ}$.\footnote{In our simulation example, targets cannot be above above 20\,km altitude target $T_1$ only exists for short ranges as a target cannot be .} Target $T_1$ represents the worst-case target parameters and $T_2$ is an optimistic case in terms of the radar resources. The green area in Figure~\ref{fig:quality function} indicates the area between $q_{k,min}$ and $q_{k,max}$ of the local utility function. If the angular estimation error is smaller than 1\,mrad then the radar would use too many resources. For angular estimation error bigger than 3\,mrad, the allocation problem would use too little resources. Clearly, at a closer range or for different target parameters, the radar can utilize the split-aperture for resource allocation. As discussed before, it is assumed that the transmit time or update frequency cannot be lowered below a certain minimum value to maintain certain track requirements such as Doppler resolution. Note that, at close range, there is a discontinuity in the lines caused by the assumption that the measurement accuracy is practically upper bounded using a maximum $\SNn$ of 40\,dB.

The resource allocation is evaluated on AMD EPYC 74F3 and the code is written in Python 3.10.11 with NumPy 1.24.3 and SciPy 1.10.1.

\subsection{Simulation Results} \label{subsec:Simulation results}

The simulation results of the split-aperture resource allocation problem with and without constrains is compared to the full-aperture resource allocation problem in this section. In the following, figures on the number of active tracks, the total utility, angular estimation error, computational time, and number of function evaluations for the same allocation problem are discussed. 

\begin{figure}%
\centering%
\input{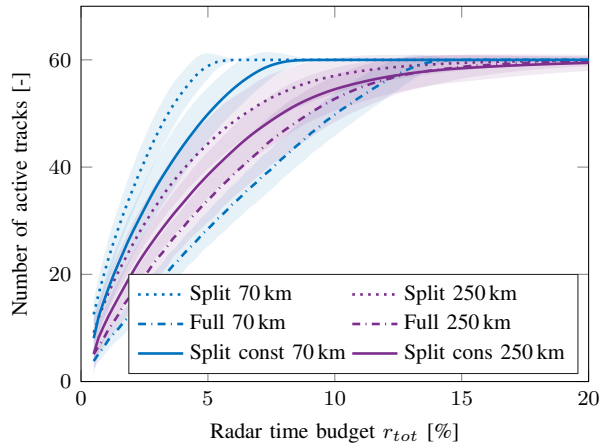}%
\caption{The number of active tracks given a radar time budget for the split-aperture and full-aperture resource problem in a scene with targets at a maximum range of 70\,km or 250\,km. The curve is the mean over $N_{MC}=100$ Monte Carlo runs and the area indicates $\pm 2\sigma$ deviation.}%
\label{fig:active tracks}%
\end{figure}

Figure~\ref{fig:active tracks} provides the number of active tracks given a radar time budget $r_{tot}$ for the three allocation problems for a scene with targets at a maximum range of either 70\,km or 250\,km. The displayed curves are the mean over $N_{MC}=100$ Monte Carlo runs and the area indicates the $\pm 2\sigma$ deviation. The figures highlights that the split-aperture concept, both constrained and unconstrained, can maintain a higher number of active tracks than the full-aperture concept for the same time budget. As expected, the constrained split-aperture concept is in between the unconstrained split-aperture and full-aperture concept. In the scene with targets at maximum 70\,km, the constrained split-aperture concept is closer to the curve of the unconstrained split-aperture concept compared to the case of maximum range of 250\,km. In this case, tasks can be fulfilled with a small aperture (less resources), leading increased number of simultaneous tasks and flexibility to schedule them on the array.

Note that the number of active tracks for the full aperture with radar time budget $0\,\%\leq r_{tot}\leq 12\,\%$ for the 70\,km case is lower than for the 250\,km case. In these cases, the SNR is not the limiting factor, but the KBS model requires an increased update frequency for fast maneuvering targets close by the radar to avoid target loss. Therefore, the individual tasks require increased radar resources leading to a decreased number of active tracks.

In Figure~\ref{fig:total utility}, the total utility of the three allocation problems is given for a radar time budget $r_{tot}$ for the three allocation problems for a scene with targets at a maximum range of either 70\,km or 250\,km. In line with the discussion of Figure~\ref{fig:active tracks}, the unconstrained split-aperture resource allocation problem achieves the highest total utility, then the constrained split-aperture concept, and followed by the full-aperture in both scenes. As expected, the total utility monotonically increases to 1 as expected.

\begin{figure}%
\centering%
\input{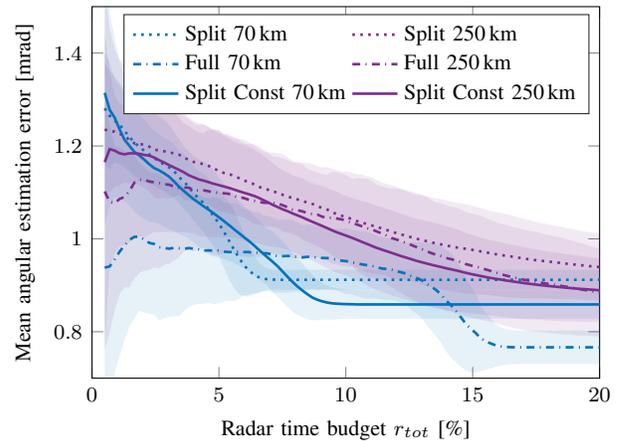}%
\caption{The mean angular estimation error given a radar time budget for the split-aperture without constrains, with constrains, and full-aperture resource problem in a scene with targets at a maximum range of 70\,km or 250\,km. The curve is the mean over $N_{MC}=100$ Monte Carlo runs and the area indicates $\pm 2\sigma$ deviation.}%
\label{fig:track sharpness}%
\end{figure}

Figure~\ref{fig:track sharpness} highlights the mean angular estimation error of the three allocation problems for a certain radar time budget $r_{tot}$ for a scene with targets at a maximum range of either 70\,km or 250\,km. The mean angular estimation error is computed for all tasks with a non-zero resource assigned. Clearly, in all resource allocation problems, the mean angular estimation error is below the $q_{k,min}=3$\,mrad, as expected. When also considering Figures~\ref{fig:active tracks} and~\ref{fig:total utility}, then it can be concluded that the angular estimation error is traded-off ($1\text{\,mrad}>q_k>3\text{\,mrad}$) to achieve a higher number of active tracks in all cases, as is the expectation of this type of resource allocation problems. If sufficient radar time budget $r_{tot}$ is available then the total utility should equal one and the mean estimation error should equal 1\,mrad. If the mean estimation error is below 1\,mrad than the resource allocation assigns more resources than strictly necessary. In such a case, either the control space $\sS$ has insufficient granularity or the minimum in the control space $s_{t,k}$ still results in an excessive aperture-time budget for that task given the environmental space $e_{t,k}$. This is most prominently visible in Figure~\ref{fig:track sharpness} for the full-aperture resource allocation with targets at a maximum range of 70\,km compared to all other cases.

\begin{figure}%
\centering%
\input{total_utility.tex}%
\caption{The total utility given a radar time budget for the split-aperture and full-aperture resource problem in a scene with targets at a maximum range of 70\,km or 250\,km. The curve is the mean over $N_{MC}=100$ Monte Carlo runs and the area indicates $\pm 2\sigma$ deviation.}%
\label{fig:total utility}%
\end{figure}

The performance of the split-aperture resource management has been demonstrated. Next, the computational aspect of the constrained split-aperture resource management will be evaluated. Figure~\ref{fig:computation time} highlights box plots with the computation time for the split-aperture without constrains and with constrains, and the full-aperture resource allocation problem in a scene with targets at a maximum range of 70\,km or 250\,km. The full-aperture resource allocation has the lowest computational overhead, as the control space $s_k$ has only two decision variables compared to the split-aperture with four. The split-aperture resource allocation without the constrains also has a low computational overhead as the tasks can be evaluated independently. For the split-aperture resource allocation with the constrains, the concave-majorant approximation algorithm must take all tasks into account. As there is an explosion in the number of possible direction for the adaptive fast traversal algorithm, this results in a significant increase in the computational overhead. Besides, it can be observed that the computational time is less for the unconstrained and constrained split-aperture resource allocation in a scene with targets at a maximum range of 250\,km. For this case, the discrete control set of the number of elements in the horizontal $N_{h,t}$ and vertical $N_{v,t}$ direction of certain tasks is limited, as a significant time-aperture budget is necessary to obtain the minimal quality requirements. 

\begin{figure}%
\centering%
%
\definecolor{mycolor1}{rgb}{0.00000,0.44706,0.74118}%
\definecolor{mycolor2}{rgb}{0.85098,0.32549,0.09804}%
\definecolor{mycolor3}{rgb}{0.92900,0.69400,0.12500}%
\definecolor{mycolor4}{rgb}{0.49400,0.18400,0.55600}%
\begin{tikzpicture}[font=\footnotesize]

\begin{axis}[%
width=0.8\columnwidth,
height=0.586\columnwidth,
at={(0\columnwidth,0\columnwidth)},
scale only axis,
boxplot/draw direction=y,
ylabel={Number of evaluations [-]},
boxplot={
    %
    draw position={1/3 + floor(\plotnumofactualtype/2) + 1/3*mod(\plotnumofactualtype,2)},
    %
    box extend=0.3,
},
x=2cm,
xtick={0,1,2,...,10},
x tick label as interval,
xticklabels={%
		{Full},%
    {Split},%
    {Split Cons},%
},
    x tick label style={
        text width=2.5cm,
        align=center
    },
ymode=log,
cycle list={{mycolor1},{mycolor4}},
legend style={at={(0.98,.02)},anchor=south east, legend cell align=left, align=left, draw=white!15!black},
]

		\addlegendimage{area legend, mycolor1,fill,fill opacity=0.2}
		\addlegendimage{area legend, mycolor4,fill,fill opacity=0.2}
		\addlegendimage{only marks, mark=*, mark options={solid, black, fill, fill opacity=1, scale=0.6}}
		\legend{70\,km, 250\,km, Outlier}

    \addplot[
    mycolor1,fill,fill opacity=0.2,
		mark=*, mark options={solid, black, fill, fill opacity=1, scale=0.6},
    boxplot prepared={
			lower whisker=290,
			lower quartile=302,
      median=308,
      upper quartile=315,
      upper whisker=334
    },
    ] coordinates {(0, 344)};

    \addplot[
    mycolor4,fill,fill opacity=0.2,
		mark=*, mark options={solid, black, fill, fill opacity=1, scale=0.6},
    boxplot prepared={
			lower whisker=374,
			lower quartile=528,
      median=596,
      upper quartile=662,
      upper whisker=789
    },
		] coordinates {};

    \addplot[
    mycolor1,fill,fill opacity=0.2,
		mark=*, mark options={solid, black, fill, fill opacity=1, scale=0.6},
    boxplot prepared={
			lower whisker=843,
			lower quartile=945,
      median=988,
      upper quartile=1032,
      upper whisker=1134
    },
		] coordinates {(0, 810) (0, 1171)};

    \addplot[
    mycolor4,fill,fill opacity=0.2,
		mark=*, mark options={solid, black, fill, fill opacity=1, scale=0.6},
    boxplot prepared={
			lower whisker=716,
			lower quartile=831.5,
      median=898.5,
      upper quartile=970.5,
      upper whisker=1112
    },
    ] coordinates {};

    \addplot[
    mycolor1,fill,fill opacity=0.2,
		mark=*, mark options={solid, black, fill, fill opacity=1, scale=0.6},
    boxplot prepared={
			lower whisker=82478,
			lower quartile=96136,
      median=99561.5,
      upper quartile=105363.5,
      upper whisker=117091
    },
    ] coordinates {(0, 75227) (0, 82187) (0, 119351) (0, 120868)};

    \addplot[
    mycolor4,fill,fill opacity=0.2,
		mark=*, mark options={solid, black, fill, fill opacity=1, scale=0.6},
    boxplot prepared={
			lower whisker=67434,
			lower quartile=95925.5,
      median=109629.5,
      upper quartile=121680.5,
      upper whisker=142583
    },
    ] coordinates {};

\end{axis}
\end{tikzpicture}%
\caption{The number of function evaluations of the $(q_k, u_k, g_k)$ pairs for the split-aperture without constrains, with constrains, and full-aperture resource allocation problem in a scene with targets at a maximum range of 70\,km or 250\,km to obtain the approximate concave-majorant. The box plot contains $N_{MC}=100$ Monte Carlo runs.}%
\label{fig:number function evaluations}%
\end{figure}
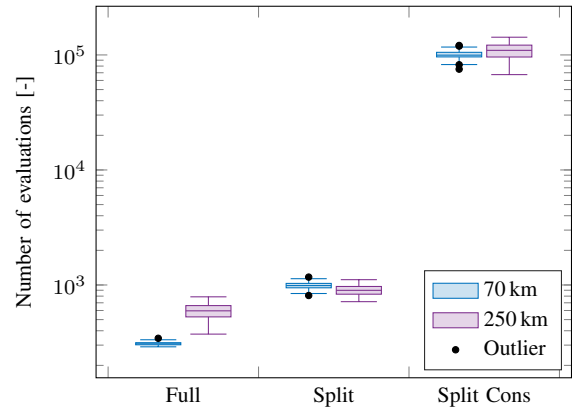

Figure~\ref{fig:number function evaluations} shows the number of evaluations of the local pairs $(q_k, u_k, g_k)$ to obtain the approximate concave-majorant. The results are in line with Figure~\ref{fig:computation time}. Clearly, to find the approximate concave-majorant of the constrained split-aperture results in a significantly higher number of function evaluations as the tasks become coupled.

The computational time and number of function evaluations for the constrained split-aperture resource allocation are not acceptable for real-time implementation. Three major steps could be taken here. First, the program could be written in, e.g., C++, for faster execution and making proper use of parallel processing during the sampling of the pair $(q_k, u_k, g_k)$. Second, a more computationally efficient version of the DBLF algorithm~\cite{Chazelle1983} can be used. Third, to avoid to approximate the full curve of the concave-majorant, the CPADS algorithm~\cite{Charlish2015a} with Algorithm~\ref{alg:AFT} could be used that is 'hot started' using the solution of the previous time-step or using the solution of the unconstrained split-aperture resource allocation. The CPADS avoids obtaining the full curve of the concave-majorant, but only searches for the point that maximizes the utility and satisfies the resource constrained.

On the other hand, the authors would like to point out that a solution is found in the control space with extremely many set-points by the AFT algorithm in a reasonable time. Moreover, to the authors' knowledge, this is the second paper on SAPA resource management and the first to consider constrained scheduling over array.

\begin{figure}%
\centering%
%
\definecolor{mycolor1}{rgb}{0.00000,0.44706,0.74118}%
\definecolor{mycolor2}{rgb}{0.85098,0.32549,0.09804}%
\definecolor{mycolor3}{rgb}{0.92900,0.69400,0.12500}%
\definecolor{mycolor4}{rgb}{0.49400,0.18400,0.55600}%
\begin{tikzpicture}[font=\footnotesize]

\begin{axis}[%
width=0.8\columnwidth,
height=0.586\columnwidth,
at={(0\columnwidth,0\columnwidth)},
scale only axis,
boxplot/draw direction=y,
ylabel={Computation time [s]},
boxplot={
    %
    draw position={1/3 + floor(\plotnumofactualtype/2) + 1/3*mod(\plotnumofactualtype,2)},
    %
    box extend=0.3,
},
x=2cm,
xtick={0,1,2,...,10},
x tick label as interval,
xticklabels={%
		{Full},%
    {Split},%
    {Split Cons},%
},
    x tick label style={
        text width=2.5cm,
        align=center
    },
ymode=log,
cycle list={{mycolor1},{mycolor4}},
legend style={at={(0.98,.02)},anchor=south east, legend cell align=left, align=left, draw=white!15!black},
]

		\addlegendimage{area legend, mycolor1,fill,fill opacity=0.2}
		\addlegendimage{area legend, mycolor4,fill,fill opacity=0.2}
		\addlegendimage{only marks, mark=*, mark options={solid, black, fill, fill opacity=1, scale=0.6}}
		\legend{70\,km, 250\,km, Outlier}

    \addplot[
    mycolor1,fill,fill opacity=0.2,
		mark=*, mark options={solid, black, fill, fill opacity=1, scale=0.6},
    boxplot prepared={
			lower whisker=0.222827434539795,
			lower quartile=0.287328243255615,
      median=0.333320975303650,
      upper quartile=0.364412903785706,
      upper whisker=0.458162069320679
    },
    ] coordinates {};

    \addplot[
    mycolor4,fill,fill opacity=0.2,
		mark=*, mark options={solid, black, fill, fill opacity=1, scale=0.6},
    boxplot prepared={
			lower whisker=0.317793846130371,
			lower quartile=0.426371693611145,
      median=0.490702390670776,
      upper quartile=0.568528890609741,
      upper whisker=0.739522218704224
    },
		] coordinates {(0, 0.8022)};

    \addplot[
    mycolor1,fill,fill opacity=0.2,
		mark=*, mark options={solid, black, fill, fill opacity=1, scale=0.6},
    boxplot prepared={
			lower whisker=2.2196,
			lower quartile=2.8185,
      median=2.9977,
      upper quartile=3.3461,
      upper whisker=3.7993
    },
		] coordinates {};

    \addplot[
    mycolor4,fill,fill opacity=0.2,
		mark=*, mark options={solid, black, fill, fill opacity=1, scale=0.6},
    boxplot prepared={
			lower whisker=1.65242433547974,
			lower quartile=1.98087453842163,
      median=2.18629300594330,
      upper quartile=2.36304795742035,
      upper whisker=2.80841588973999
    },
    ] coordinates {(0,3.1453)};

    \addplot[
    mycolor1,fill,fill opacity=0.2,
		mark=*, mark options={solid, black, fill, fill opacity=1, scale=0.6},
    boxplot prepared={
			lower whisker=5602.79036664963,
			lower quartile=7057.69372022152,
      median=7600.62037706375,
      upper quartile=8166.23728013039,
      upper whisker=9664.73234105110
    },
    ] coordinates {(0, 4732.58608818054) (0, 10534.2305300236)};

    \addplot[
    mycolor4,fill,fill opacity=0.2,
		mark=*, mark options={solid, black, fill, fill opacity=1, scale=0.6},
    boxplot prepared={
			lower whisker=2599.30126905441,
			lower quartile=4054.77885138989,
      median=4747.28614330292,
      upper quartile=5402.00860619545,
      upper whisker=6811.15834164619
    },
    ] coordinates {(0, 2004.53194618225)};

\end{axis}
\end{tikzpicture}%
\caption{The computation time for the split-aperture without constrains, with constrains, and full-aperture resource allocation problem in a scene with targets at a maximum range of 70\,km or 250\,km. The box plot contains $N_{MC}=100$ Monte Carlo runs.}%
\label{fig:computation time}%
\end{figure}
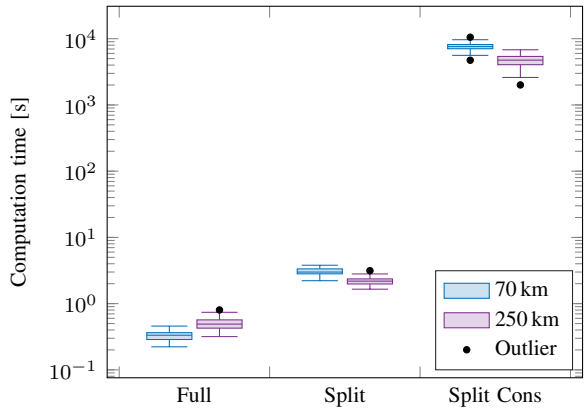

\section{\uppercase{Conclusions}} \label{sec:conclusions}
In this paper, radar resource management has been introduced for the split-aperture phased array concept and the added benefit of the SAPA concept for active tracking tasks has been demonstrated. In particular, the Q-RAM framework has been used with a modification of the KBS model to formulate SAPA resource management for active tracking tasks. To obtain a solution of the allocation problem, the introduced adaptive fast traversal algorithm has been combined with a three dimensional strip packing algorithm. The adaptive fast traversal algorithm can efficiently handle discontinuities in the concave-majorant approximation coming from resource allocation over the array. It has been shown by a simulation example that the SAPA concept can significantly increase the number of active tracks of a multifunction radar system compared to scheduling tasks sequentially. Hence, the added flexibility of the SAPA concept can significantly increase efficiency of multifunction phased array radar systems.

Topics for future research are to reduce the computational effort of the SAPA resource management and to included tracker independent performance measures similar to the (Bayesian) Cram\'{e}r-Rao lower bound. In addition, it is planned to study the impact of the parameters $\alpha_1$, $n_1$, $n_2$, $n_3$ of the adaptive fast traversal algorithm on the performance of approximating the concave-majorant.

\bibliographystyle{IEEEtaes_DOI}

\bibliography{bibliography}

\end{document}